\documentclass[showpacs,preprintnumbers,amssymb,amsmath]{revtex4}

\usepackage[english]{babel}

\begin{document}

\bibliographystyle{apsrev}

\title{ Equivalence of Markov's Symbolic Sequences to Two-Sided Chains}

\author{S. S. Apostolov, Z. A. Mayzelis}
\affiliation{V. N. Karazin Kharkov National University, 4 Svoboda
Sq., Kharkov 61077, Ukraine}

\author{O. V. Usatenko, V. A. Yampol'skii
\footnote[1]{yam@ire.kharkov.ua} }
\affiliation{A. Ya. Usikov Institute for Radiophysics and Electronics \\
Ukrainian Academy of Science, 12 Proskura Street, 61085 Kharkov,
Ukraine}

\begin{abstract}
A new object of the probability theory, two-sided chain of events
(symbols), is introduced. A theory of multi-steps Markov chains
with long-range memory, proposed earlier in Phys. Rev. E
\textbf{68}, 06117 (2003), is developed and used to establish the
correspondence between these chains and two-sided ones. The Markov
chain is proved to be statistically equivalent to the definite
two-sided one and vice versa. The results obtained for the binary
chains are generalized to the chains taking on the arbitrary
number of states.
\end{abstract}

\maketitle

\section{Introduction}
\label{I}

The problem of long-range correlated random symbolic systems (LRCS)
has been under study for a long time in many areas of contemporary
physics~\cite{bul,sok,bun,yan,maj,halvin},
biology~\cite{vossDNA,stan,buld,prov,yul,hao},
economics~\cite{stan,mant,zhang},
linguistics~\cite{schen,kant,kokol,ebeling,uyakm}, etc.

Among the ways to get a correct insight into the nature of
correlations of complex dynamic systems the use of the
\emph{multi-step Markov} chains is one of the most important because
they give a possibility to construct a random sequence with
necessary correlated properties in the most natural way. This was
demonstrated in Ref.~\cite{uya}, where the concept of Markov chain
with the \emph{step-wise memory function}, which consist in
coordinate independence of the conditional probability, was
introduced. The concept of additive chains turned out to be very
useful due to the ability to evaluate the binary correlation
function of the chain through the memory function (see for the
details Refs.~\cite{mel,AllMemStepCor}). The correlation properties
of some dynamic systems (coarse-grained sequences of the
\emph{Eukarya's DNA and dictionaries}) can be well described by this
model~\cite{uya}.

Another important reason for the study of Markov chains is its
application to the various physical objects~\cite{tsal,abe,den},
e.g., to the Ising chains of spins. The problem of thermodynamics
description of the Ising chains with long-range spin interaction is
opened even for the 1D case. However, the association of such
systems with the Markov chains can shed light on the non-extensive
thermodynamics of the LRCS.

Multi-step Markov chains are characterized by the probability that
each symbol of the sequence takes on the definite value under
condition that some \textit{previous} symbols are fixed. This chains
can be easily constructed by the consequent generation using
prescribed conditional probability function. Besides, the
statistical properties of Markov chains can be determined in some
simple cases. At the same time, there is another class of correlated
sequences, the so-called two-sided chains. They are determined by
the probability that each symbol of the sequence takes on the
definite value under the condition that some symbols \emph{at the
both sides} of the chosen symbol are fixed. An example of systems
with such property is the above-mentioned Ising chain. But the
approach, used for the finding of Markov chains properties (the
probability of concrete ''word'' occurring, the correlation
functions, and so on) unfortunately cannot be used in this case. In
this paper we prove, that such mathematical objects, determined in
the Sec.~(\ref{def}) as two-sided chains, are the Markov chains. So,
the statistical properties of Markov chains and the method of their
constructing can be used for the studying the two-sided chains.

The paper is organized as follows. In the first Section we give
the definition of Markov and two-sided chains. The next Section is
devoted to the proof of the main statement: the first Subsection
contains the proof of the direct statement, that every binary
Markov chain is in the same time the binary two-sided one; the
second Subsection shows, that the classes of these two chains
coincide. Finally, in the last Subsection we generalize this
results to the case of non-binary chains.

\section{Basic notions}

\subsection{General definitions} \label{def}

Let us determine the \textit{$N$-step Markov} chain. This is a
sequence of random variables~$a_i$, $i=-M,-M+1,\ldots,M$ ($M\gg N$),
referred to as the \textit{symbols}, which have the following
property: the probability of symbol~$a_i$ to have a certain value
under the condition that the values of all \textbf{previous} symbols
are fixed depends on the values of~$N$ previous symbols only,
\begin{equation}\label{def_mark}
 P(a_i=a|\ldots,a_{i-2},a_{i-1})=
 P(a_i=a|a_{i-N},\ldots,a_{i-2},a_{i-1}).
\end{equation}
Such defined chain is a \textit{stationary} one, because the
conditional probability does not depend explicitly on $i$, i.e.,
does not depend on the position of symbols
$a_{i-N},\ldots,a_{i-1},a_{i}$ in the chain and depends on the
values of the symbols and their positional relationship only.

The chain under consideration is defined for arbitrary but finite
length $M$. Nevertheless, all results that will be obtained do not
depend on $M$. Thus, they are correct in the infinite limit provided
that the conditional probability function is fixed.

In different mathematical and physical problems we confront with
the sequences for which the probability of symbol~$a_i$ to have
certain value, under the condition that the values of the
\textbf{rest} of symbols are fixed, depends on a value of~$N$
previous and~$N$ next symbols only,
\[P(a_i=a|\ldots,a_{i-2},a_{i-1},a_{i+1},a_{i+2},\ldots)=
\quad\quad\phantom{stas}
\]
\begin{equation}\label{def_2s}
\phantom{stas}\quad\quad
=P(a_i=a|a_{i-N},\ldots,a_{i-1},a_{i+1},\ldots,a_{i+N}).
\end{equation}
Let us name these chains as~\textit{$N$-two-sided} ones. By the
same reason as above, see Eq.~(\ref{def_mark}), this chain is a
stationary one.

An important class of the random sequences is the \textit{binary}
chains. If each symbol~$a_i$ of the chain can take on only two
values, $s_0$~and~$s_1$, then we refer to this chain as a binary.
It is convenient to change a value of~$a_i$ to~$0$~and~$1$ using a
linear transformation,
\[ a_i:=\frac{a_i-s_0}{s_1-s_0}. \]

Now, we will describe the ways of the constructing of the defined
chains.

\subsection{Constructing of the chains}

The Markov chain defined in such way is simple for numerical
simulations. There are two basic approaches for this. In both of
them we find successively the each next generated symbol by $N$
previous ones. But these approaches differ in the method of
constructing for the first \textit{$N$-word}, the set of $N$
sequent symbols.

For first approach one needs to calculate in addition some
conditional probabilities. They can be found from the compatibility
equation for the conditional probabilities:
\begin{equation}\label{soglas}
P(a_i=a|a_{i-k},\ldots,a_{i-1})=\frac{\sum\limits_{a_{i-N}}\!\ldots\!\!
\sum\limits_{a_{i-k-1}}\!\! P(a_i=a|a_{i-N},\ldots,a_{i-1})
P(a_{i-N},\ldots,a_{i-1}) } {\sum\limits_{a_{i-N}}\!\ldots\!\!
\sum\limits_{a_{i-k-1}}\!\!P(a_{i-N},\ldots,a_{i-1})}.
\end{equation}
Here $k=0,\ldots,N-1$ and the sign $\sum\limits_{a_{j}}$ means
summation (or integration) over all possible values of symbol $a_j$.
The probabilities of $N$-words occurring,
$P(a_{i-N},\ldots,a_{i-1})$, should be obtained from the following
linear system,
\begin{equation}\label{for_word}
\left\{\begin{array}{l} P(a_1,a_2,\ldots, a_N)=
\sum\limits_{a_0}P(a_N|a_0,a_1,\ldots, a_{N-1})P(a_0,a_1,\ldots,
a_{N-1}),\\
\sum\limits_{a_1}\ldots \sum\limits_{a_N} P(a_1,a_2,\ldots, a_N)=1.
\end{array}
\right.
\end{equation}
Using Eq.~\eqref{soglas} we can construct the first $N$ symbols
consequently generating them in accordence to the following
conditional probabilities:
\[P(a_1),\,P(a_2|a_1),\,P(a_3|a_1,a_2),\ldots,
P(a_N|a_1,\dots,a_{N-1}).\]

The second approach is based on the random choice of $N$-word. The
second approach is simpler than the first one because it does not
make the calculation of additional probabilities. But it does not
allows to get the stationary chain, as it is possible in the first
method. For generation of the same chain using the second approach
we must construct as many symbols as one needs to get the
stationary chain (the initial part of the chain should be
removed).

There is no simple method for two-sided chains constructing. The
best known and simple approach is Metropolis' algorithm, but it
needs much more operations than constructing  of the Markov chain.
Therefore it is very important to prove the equivalence of the
Markov and two-sided chains.

\section{
equivalence of the Markov and two-sided chains} \label{prm}

In this section we prove an equivalence of two random sequences, the
Markov and two-sided chains. The proof is produced for a binary
chain, but it can be directly generalized for arbitrary chains (see
Subsection~\ref{sec_non_bi} for details). This proof requires some
formulas for a conditional probability. Its definition is
\begin{equation}\label{def_uv}
P(A|B)=\frac{P(A,B)}{P(B)}.
\end{equation}
Here and below comma between the symbols-events means that both of
these events occur simultaneously, it is a product of two events,
$(A,B)=A\bigcap B$. Using evident equation,
\begin{equation}\label{perehod}
P(A,B|C)=P(A|B,C)P(B|C),
\end{equation}
the following formula can be easily obtained,
\begin{equation}\label{for_dsv}
P(A|B,C)=\frac{P(A,B|C)}{P(A,B|C)+P(\overline{A},B|C)},
\end{equation}
where $\overline{A}$ is an event opposite to~$A$.

\subsection{From the Markov to two-sided chain\label{m_to2s}}

Let us demonstrate that a Markov chain is a two-sided one. For this
purpose using Eq.~(\ref{for_dsv}) we rewrite the probability for
symbol~$a_i$ to be equal unity, under the condition that the values
of the \textbf{rest} of symbols are fixed, in following form:
\begin{equation}\label{dsv1}
P(a_i=1|A_i^-,A_i^+)=\displaystyle\frac{P(a_i=1,A_i^+|A_i^-)}
{P(a_i=1,A_i^+|A_i^-)+P(a_i=0,A_i^+|A_i^-)}.
\end{equation}
 where~$A_i^-=(\ldots,a_{i-2},a_{i-1})$
and~$A_i^+=(a_{i+1},a_{i+2},\ldots)$.

To obtain the value of~$P(a_i=1,A_i^+|A_i^-)$ one needs to use
Eq.~(\ref{perehod}) many times to express $P(.|.)$ as the
product:
\[
P(a_i=1,A_i^+|A_i^-)=P(a_i=1|A_i^-)
P(a_{i+1},A_{i+1}^+|a_i=1,A_i^-)=
 \]
\[
=P(a_i=1|A_i^-)P(a_{i+1}|a_i=1,A_i^-)
P(a_{i+2},A_{i+2}^+|A_i^-,a_i=1,a_{i+1})=\ldots
 \]
\begin{equation}\label{dsv2}
\ldots=\prod\limits_{{r=i\atop a_i=1}}^{M} P(a_{r}|A_{r}^-).
\end{equation}
However the chain under consideration is the~$N$-step Markov one
and, according to definition~(\ref{def_mark}), the probability of
symbol~$a_i$, under the condition that the  values of all previous
symbols are fixed, depends on the values of~$N$ previous symbols
only. So, the factors of the product for~$r>i+N$ in Eq.~(\ref{dsv2})
do not depend on~$a_i$. Substituting expression~(\ref{dsv2})
for~$P(a_i=1,A_i^+|A_i^-)$ into Eq.~(\ref{dsv1}) we derive the
following equation,
\begin{equation}\label{dsv}
P(a_i=1|T_i^-,T_i^+)=\displaystyle\frac{ \prod\limits_{{r=0\atop
a_i=1}}^N P(a_{i+r}|T^-_{i+r})} {\prod\limits_{{r=0\atop a_i=1}}^N
P(a_{i+r}|T^-_{i+r})+\prod\limits_{{r=0\atop a_i=0}}^N
P(a_{i+r}|T^-_{i+r})}.
\end{equation}
Here $T^-_{j}=(a_{j-N},\ldots,a_{j-1})$ and
$T^+_{j}=(a_{j+1},\ldots,a_{j+N})$ are previous and next words of
the length~$N$ with respect to symbol~$a_j$.

Equation~(\ref{dsv}) is the fundamental relation for association of
Markov and two-sided chains. One can see from it that the
probability of symbol~$a_i$ under the condition of fixed values of
the \textbf{rest} of symbols is determined only by two words of the
length~$N$, $T^-_{i}$ and~$T^+_{i}$. So, according to
definition~(\ref{def_2s}), the Markov chain is the two-sided one,
quod erat demonstrandum.

\subsection{From two-sided to the Markov chain\label{2s_tom}}

Now we prove the opposite statement: the two-sided chain is a Markov
one. I.e., we prove that the probability of the symbol~$a_i$ to be
equal to unity, under the condition that all \textbf{previous}
symbols are fixed, depends on the values of~$N$ previous symbols
only. Thereto, let us take two sets of symbols~$A'$ and~$A''$ which
are two variants of the word~$A_{i-N}^-$ and differ only by one
symbol~$a_{i-N-k}$ at arbitrary value of~$k>0$.

Using definition of the conditional probability~(\ref{def_uv}) we
obtain
\[P(a_i|A',T^-_{i})=\frac{P(A',T^-_{i},a_i)}{P(A',T^-_{i})}=
\frac{P(a'_{i-N-k}|\tilde{A},T^-_{i},a_i)
P(\tilde{A},T^-_{i},a_i)}
{P(a'_{i-N-k}|\tilde{A},T^-_{i})P(\tilde{A},T^-_{i})}
=\]\[=\frac{P(a'_{i-N-k}|\tilde{A},T^-_{i},a_i)
}{P(a'_{i-N-k}|\tilde{A},T^-_{i})}P(a_i|\tilde{A},T^-_{i}),
\]
where~$\tilde{A}$ is a set of symbols~$A'$ (or~$A''$) except for
symbol~$a_{i-N-k}$. However, according to the definition of
two-sided chain, conditional
probability~$P(a'_{i-N-k}|\tilde{A},T^-_{i},a_i)$ does not depend on
symbol~$a_i$ since the latter is situated at a distance more
than~$N$ from~$a_{i-N-k}$. Hence one gets
\[P(a_i|A',T^-_{i})=P(a_i|\tilde{A},T^-_{i})=
P(a_i|A'',T^-_{i}).\]

So, we find that probability~$P(a_i|A_{i}^-)$ takes on the same
value for any arbitrary word~$A_{i-N}^-$. We conclude that the
probability does not depend on~$A_{i-N}^-$. Thus we attest ourselves
that
\[P(a_i|A^-_{i})=P(a_i|T^-_{i}).\]
In other words, according to definition~(\ref{def_mark}), the
two-sided chain is a Markov one, quod erat demonstrandum.

It should be emphasized that every two-sided chain is equivalent to
the \textbf{single} Markov one though it is not evident because of
the non-linear structure of Eq.~(\ref{dsv}). Using trivial
expression of Eq.~(\ref{def_uv}),
\begin{equation}\label{for_mark}
P(a_i|T^-_{i})=\frac{P(T^-_{i},a_i)}{P(T^-_{i})},
\end{equation}
one can easily make sure that a single chain cannot have two
different conditional probabilities. The matter is that the
probabilities of $N$- and $(N+1)$-words occurring determines
uniquely the conditional probability according to
Eq.~(\ref{for_mark}). Hence, for the chain under study the Markov
conditional probability is determined uniquely.

\subsection{\label{sec_non_bi}The case of non-binary chain}

The results obtained in previous Secs.~\ref{m_to2s} and~\ref{2s_tom}
can be generalized to non-binary chains. And we can develop the
similar proof and get the following equation connecting the
conditional probability functions,
\begin{equation}\label{for_nbc}
P(a_i=a|T^-_{i},T^+_{i}) = \displaystyle\frac{\prod\limits_{r=0\atop
a_i=a}^N P(a_{i+r}|T^-_{i+r})} { \sum\limits_{\xi\in
\mathcal{A}}\prod\limits_{r=0\atop a_i=\xi}^N
P(a_{i+r}|T^-_{i+r})},
\end{equation}
that is analogue of Eq.~(\ref{dsv}).

In this formula we used the following notations:

-- if symbol~$a$ takes on the finite set of values~$\mathcal{A}$
then we use the conditional probabilities $P(a|\ldots)$;

-- if symbol~$a$ takes on the continuous set of
values~$\mathcal{A}$ then we used conditional probability
density~$P(a|\ldots)$ and sign~$\sum\limits_{\xi\in\mathcal{A}}$
means $\int\limits_\mathcal{A}d\xi$.

Thus, the equivalence between the N-two-sided and N-step Markov
chains is proved for non-binary chains also. We found the very
important formula for the conversion the Markov's conditional
probability to the two-sided one and inversely. This method can be
used for numerical and analytic calculations of the conditional
probabilities.

\section{Conclusion}

Thus, we proved that the classes of the ``one-sided'' Markov
chains and two-sided ones coincide. The obtained relationship
between the conditional probabilities (or its densities in the
case of continuous distribution of values taking on by the
elements of the chains) allows to construct numerically the Markov
chain possessing the same statistical properties as the initial
two-sided one. So, two-sided sequence can be easily reproduced
numerically with conservation of \emph{all} statistical properties
but not binary correlation function as it was done in the
papers~\cite{mel,AllMemStepCor}.

Besides, found Eq.~\eqref{for_nbc} allows to use results of
analytical studies of Markov chains (for example,
see~\cite{AllMemStepCor}) for the two-sided sequences. This can be
very useful for the study of physical systems. The example is the
Ising chain, that is the two-sided one.

\end{document}